# *Continuum Photonic Axion Insulators*

*Mário G. Silveirinha*[(1)1]

[(1)] *University of Lisbon–Instituto Superior Técnico and Instituto de Telecomunicações, Avenida Rovisco Pais, 1, 1049-001 Lisboa, Portugal*

**Abstract**

Here, I study the conditions for realizing an axion topology in a homogeneous inversion-symmetric electromagnetic continuum, in the absence of a crystalline lattice. I introduce a continuum inversion invariant expressed solely in terms of the spectral properties of the material response in the zero- and infinite-wave-vector limits. Surprisingly, it is found that standard gyrotropic-type material responses are strongly constrained and, under fairly general conditions, cannot open an axion gap within first-order perturbation theory. Based on an analogy between the effective response of crystalline photonic axion insulators and the optical response of centrosymmetric altermagnets, a class of spatially dispersive odd magnetoelectric couplings that can open an axion gap is identified. These findings suggest an interesting hierarchy of nonreciprocal mechanisms in electromagnetic continua: while uniform gyrotropy relates to Chern phases, odd magnetoelectric coupling relates to axion phases.

---

[1] To whom correspondence should be addressed: E-mail: mario.silveirinha@tecnico.ulisboa.pt

# I. Introduction

Topological photonics has established an explicit connection between the geometry of electromagnetic eigenstates in momentum space and the emergence of robust boundary states. The most recognizable example of this paradigm in nonreciprocal electromagnetics is the Chern topology, for which the bulk bands are characterized by a topological invariant, and the difference between the bulk topological indices (gap Chern numbers) determines the net number of unidirectional edge states crossing a common band gap [1-6].

These ideas were originally developed for photonic crystals, for which the discrete translational symmetry naturally restricts the wave vector to a compact Brillouin zone. It was later recognized that a similar topological classification can be introduced for electromagnetic continua with no intrinsic periodicity [7]. In this case, the wave-vector space is not a Brillouin zone but an unbounded Euclidean space. For two-dimensional Chern-type problems, a well-defined topological classification requires a suitable behavior of the eigenstates in the large-wave-vector limit, such that the $(k_x, k_y)$ plane can be compactified into the Riemann sphere [7]. This condition can be enforced through a wave-vector cutoff of the nonreciprocal response or can follow from a physically motivated nonlocal material model [7-12]. The topology of a continuum is then sensitive to the ultraviolet behavior of the material response, i.e., it is not uniquely determined by the local (long-wavelength) constitutive model [12, 13]. In fact, a key point is that the proper modeling of the short-wavelength response is essential to avoid singular field localization and violations of bulk-edge correspondence [12, 14-17]. The topological theory of continuous electromagnetic systems has enabled the characterization of Chern phases and the associated edge states of magnetized plasmas, ferrites, and related magnetoplasmonic systems [7-12, 18-19]. Topological phenomena associated with Berry monopoles and Weyl points have also been studied [20, 21]. Similar topological considerations arise in other continuum wave systems. In particular, rotating and odd-viscous fluids provide important examples of continuum topological waves [22-25].

The Chern class does not exhaust the possible topological phases of nonreciprocal systems. A distinct possibility is provided by axion insulators, whose topological characterization is intrinsically three dimensional. The nomenclature of such systems originates from their connection with axion electrodynamics [26]. Following the theoretical development of three-dimensional electronic topological insulators [27, 28], it was recognized that nontrivial phases can exhibit a quantized isotropic magnetoelectric response characterized by a topological invariant $\theta$ [29, 30]. The parameter $\theta$ can be expressed in terms of a three-dimensional Chern-Simons integral of the Berry connection [30]. For a time-reversal-invariant electronic insulator, symmetry restricts $\theta$ to 0 or $\pi$, with the strong topological phase corresponding to $\pi$. Importantly, $\theta$ remains quantized even when time-reversal symmetry is broken, provided inversion is preserved and the bulk Chern indices vanish. This observation led to the concept of the magnetic axion insulator [31-33]. For a nontrivial phase with $\theta = \pi$ and vanishing bulk Chern indices, a gapped surface exhibits a half-quantized Hall response, modulo an integer contribution. Interfaces between surfaces with different Hall responses may consequently support chiral hinge states.

Recently, the axion-insulator concept has been transferred to photonics. Specifically, Devescovi *et al.* showed that a relative axion phase can be engineered in three-dimensional gyrotropic photonic crystals, with inversion-symmetric domain walls supporting chiral hinge states [34]. More recently, these ideas have been experimentally demonstrated [35, 36]. The experiments demonstrated half-quantized surface Chern numbers and chiral hinge transport expected from the axion phase.

These developments have so far concerned periodic photonic structures. By contrast, a corresponding theory of axion topology for continuous electromagnetic materials, with no underlying lattice, is still lacking. Such a theory can be useful to pinpoint more clearly the physical ingredients and mechanisms that underlie the topology of these systems. Here, I show that electromagnetic continua can indeed support nontrivial axion phases. To this end, I develop a continuum formulation of the three-dimensional axion invariant and identify the conditions on the material response that make this invariant well defined. In particular, the analysis clarifies the role of the large-wave-vector response and identifies an interesting physical connection between these systems and centrosymmetric altermagnets.

## II. Axion invariant

### *A. Construction of the axion invariant*

The construction of the axion invariant is inspired by the formalism of Turner *et al.* [33]. For periodic systems with inversion symmetry and vanishing bulk Chern numbers, the authors of Ref. [33] showed that the Chern-Simons invariant can be determined solely from the inversion parities of the eigenstates at the inversion-invariant points of the Brillouin zone. Specifically, for a fully gapped system, the total number of odd-parity eigenstates below the gap, summed over all inversion-invariant momenta, must be even. Furthermore, the Chern-Simons invariant is nontrivial when this number is not divisible by four.

To extend these ideas to an inversion-symmetric electromagnetic continuum, first I note that, for a compactified wave-vector space, the only points invariant under inversion are the origin, ($\mathbf{k}=0$), and the point at infinity ($\mathbf{k}=\infty$). This suggests that the axion invariant ($\theta=0$ or $\theta=\pi$) should be defined as

$$2\frac{\theta}{\pi}=n_-\left(\Gamma\right)+n_-\left(\infty\right) \quad (\mathrm{mod}\,4), \tag{1}$$

where $n_-$ is the number of eigenstates with odd parity below the relevant gap, evaluated at the relevant point of the $\mathbf{k}$-space.

In the absence of a well-defined compactification, the asymptotic behavior of the eigenfunctions at $\mathbf{k}=\infty$ typically depends on the direction of the wave vector, which renders $n_-\left(\infty\right)$ ill-defined. This difficulty can be circumvented using the following rationale.

To begin with, the analysis is restricted to the class of inversion-symmetric materials with a response $\mathbf{M}\left(\omega,\mathbf{k}\right)$ that asymptotically is reciprocal:

$$\mathbf{M}_\infty\left(\omega\right)=\lim_{k\to\infty}\mathbf{M}\left(\omega,\mathbf{k}\right) \quad \text{is reciprocal.} \tag{2}$$

Here, $\mathbf{M}$ represents a 6×6 matrix that includes the permittivity, permeability and magnetoelectric tensors of the material [7, 37]:

$$\mathbf{M}\left(\omega,\mathbf{k}\right)=\begin{pmatrix}\boldsymbol{\varepsilon}\left(\omega,\mathbf{k}\right) & \boldsymbol{\xi}\left(\omega,\mathbf{k}\right)\\ \boldsymbol{\zeta}\left(\omega,\mathbf{k}\right) & \boldsymbol{\mu}\left(\omega,\mathbf{k}\right)\end{pmatrix}. \tag{3}$$

Since $\mathbf{M}_\infty\left(\omega\right)$ is itself inversion-symmetric, it necessarily has the form:

$$\mathbf{M}_\infty\left(\omega\right)=\begin{pmatrix}\boldsymbol{\varepsilon}_\infty\left(\omega\right) & 0\\ 0 & \boldsymbol{\mu}_\infty\left(\omega\right)\end{pmatrix}. \tag{4}$$

For a reciprocal material, $\mathbf{M}_\infty$ must be a symmetric matrix. It is assumed that this reciprocal reference shares a common gap with the original system.

In electromagnetic linear conservative systems, reciprocity correlates with a time-reversal operator that satisfies $\mathcal{T}^2 = \mathbf{1}$ [38]. Thus, reciprocal systems fall into the symmetry class AI, which in three dimensions has no strong topological-insulator phase and therefore no strong axion phase [39]. As the proposed continuum construction is intended to mimic the properties of crystalline systems, it is natural to require that a reciprocal continuum obeys the same symmetry classification as its periodic counterpart, and thus belongs to the trivial axion sector. Accordingly, it is assumed that the system described by $\mathbf{M}_\infty$ satisfies:

$$0 = n_-^\infty(\Gamma) + n_-^\infty(\infty) \quad (\mathrm{mod}\, 4)\,. \tag{5}$$

Furthermore, since $\mathbf{M}_\infty(\omega)$ and $\mathbf{M}(\omega, \mathbf{k})$ have the same asymptotic response, the contribution from infinity cancels in the relative invariant. Hence, for all the relevant purposes, one can take $n_-(\infty) = n_-^\infty(\infty)$. Combining Eqs. (1) and (5), one then arrives at the following axion symmetry indicator:

$$2\frac{\theta}{\pi} = n_-(\Gamma) - n_-^\infty(\Gamma) \quad (\mathrm{mod}\, 4)\,. \tag{6}$$

As seen, the axion invariant is determined by the difference between the number of eigenstates that have odd parity at the origin (below the relevant gap) in the considered system and in its reciprocal asymptotic form. A connection between eigenstate parity and a $\mathbb{Z}_2$ topological invariant was previously noted for time-reversal-invariant photonic continua [40]. In the next section, it is shown that the above quantity can be directly expressed in terms of the permittivity response of the material.

In general, the material dispersion is most naturally described in an augmented formulation of Maxwell equations [7]. In the augmented problem, the system is described by some operator $\hat{H}_{\mathbf{k}}$, which depends on the internal degrees of freedom of the material [7]. The identification of $\mathbf{M}_\infty(\omega)$ with the asymptotic system response should be understood in this framework. Specifically, it is assumed that, for large $k$, the material sector of the augmented operator approaches an asymptotic form, while the Maxwell sector retains its usual $\mathbf{k}$-dependence. The corresponding reference problem $\hat{H}_{\infty,\mathbf{k}}$ is defined by this asymptotic material sector for all $\mathbf{k}$ [7]. Its reduced electromagnetic response is $\mathbf{M}_\infty(\omega)$. Thus, $n_-(\Gamma)$ and $n_-^\infty(\Gamma)$ are determined by the eigenstates of $\hat{H}_{\mathbf{k}}$ and $\hat{H}_{\infty,\mathbf{k}}$, respectively.

It is useful to note that in a continuum subject to the constraint (2) a full gap automatically implies vanishing Chern invariants. Indeed, the gap Chern number evaluated at any 2D cut of the three-dimensional wave-vector space (e.g., a plane passing through the origin) can be continuously deformed, without closing the gap, into the Chern number of another parallel plane placed at $k = \infty$. However, as the system approaches a fully gapped reciprocal system for $k \to \infty$, the gap Chern number evaluated at the asymptotic plane is necessarily trivial. Thereby, the gap Chern number evaluated at any other 2D cut of the material has necessarily the same property.

*B. Link between the axion invariant and the positive inertia of the permittivity matrix*

For plane wave solutions, the Maxwell equations in a dispersive homogeneous material with response $\mathbf{M}(\omega, \mathbf{k})$ can be written in the spectral domain as:

$$\mathcal{D}(\mathbf{k}) \cdot \mathbf{f}_{\mathbf{k}} = \omega \mathbf{M}(\omega, \mathbf{k}) \cdot \mathbf{f}_{\mathbf{k}}, \quad \text{with} \quad \mathcal{D}(\mathbf{k}) = \begin{pmatrix} 0 & -\mathbf{k} \times \mathbf{1} \\ \mathbf{k} \times \mathbf{1} & 0 \end{pmatrix}. \tag{7}$$

Here, $\mathbf{f}_\mathbf{k} = \begin{pmatrix}\mathbf{E}_\mathbf{k} & \mathbf{H}_\mathbf{k}\end{pmatrix}^T$ is a six-vector of the electromagnetic fields. Clearly, $n_-(\Gamma)$ is determined by the eigenstates of (7), namely the number of eigenstates with odd parity at $\mathbf{k}=0$. Since, by assumption, the material is inversion symmetric, it is necessary that $\mathbf{M}(\omega,\mathbf{k}) = \boldsymbol{\sigma}_z \cdot \mathbf{M}(\omega,-\mathbf{k}) \cdot \boldsymbol{\sigma}_z$, where $\boldsymbol{\sigma}_z = \begin{pmatrix}\mathbf{1}_{3\times3} & 0 \\ 0 & -\mathbf{1}_{3\times3}\end{pmatrix}$ is a generalized Pauli matrix. Here, I take into account that the electric (magnetic) field is odd (even) under a space inversion. In particular, it follows that at the origin, the magnetoelectric response must be trivial (here, it is assumed that the material response is analytic at $\mathbf{k}=0$):

$$\mathbf{M}(\omega,\mathbf{k}=0) = \begin{pmatrix}\boldsymbol{\varepsilon}(\omega,\mathbf{k}=0) & 0 \\ 0 & \boldsymbol{\mu}(\omega,\mathbf{k}=0)\end{pmatrix}. \tag{8}$$

Hence, since $\mathcal{D}(\mathbf{k}=0)=0$, one sees that the electric and magnetic responses become decoupled in the $\mathbf{k}=0$ limit. Moreover, the eigenstates with odd parity take the form $\mathbf{f} = \begin{pmatrix}\mathbf{E} & 0\end{pmatrix}^T$, whereas the eigenstates with even parity take the form $\mathbf{f} = \begin{pmatrix}0 & \mathbf{H}\end{pmatrix}^T$.

As $n_-(\Gamma)$ counts the number of eigenstates with odd parity, it follows that it can be identified with the number of roots of $\det\left(\omega\boldsymbol{\varepsilon}(\omega,\mathbf{k}=0)\right)=0$ with $\omega<\omega_\mathrm{g}$, where $\omega_\mathrm{g}$ is a frequency in the relevant gap. The roots are counted with their mode degeneracy and include negative frequencies.

Now, an important observation is that for passive lossless systems $\partial_\omega\left[\omega\mathbf{M}(\omega,\mathbf{k})\right]$ is required to be a positive definite matrix, as it determines the metric that controls the energy stored in the material [7]. Consequently, $\partial_\omega\left[\omega\boldsymbol{\varepsilon}(\omega,\mathbf{k}=0)\right]$ is also positive definite, which implies that the eigenvalues of $\omega\boldsymbol{\varepsilon}(\omega,\mathbf{k}=0)$ are strictly increasing functions of frequency between the poles of the material response. Clearly, the roots of $\det\left(\omega\boldsymbol{\varepsilon}\left(\omega,\mathbf{k}=0\right)\right)=0$ occur when one of the eigenvalues of $\omega\boldsymbol{\varepsilon}\left(\omega,\mathbf{k}=0\right)$ crosses zero. As the eigenvalues are strictly increasing functions of frequency, the zeros of the eigenvalues necessarily alternate with their poles. Hence, $n_-\left(\Gamma\right)$ must coincide with the total number of poles of $\det\left(\omega\boldsymbol{\varepsilon}\left(\omega,\mathbf{k}=0\right)\right)$ with $\omega<\omega_{\mathrm{g}}$ (again counted with their degeneracy) plus the number of positive eigenvalues of $\omega_{\mathrm{g}}\boldsymbol{\varepsilon}\left(\omega_{\mathrm{g}},\mathbf{k}=0\right)$:

$$n_-\left(\Gamma\right)=N_{\mathrm{p}}\Big|_{\omega<\omega_{\mathrm{g}}}+\mathrm{pos}\left(\omega_{\mathrm{g}}\boldsymbol{\varepsilon}\left(\omega_{\mathrm{g}},\mathbf{0}\right)\right). \tag{9}$$

The number of positive eigenvalues of a matrix $\mathbf{A}$ is known as the positive inertia of the matrix ($\mathrm{pos}\left(\mathbf{A}\right)$). Note that a positive eigenvalue indicates that an additional zero of $\det\left(\omega\boldsymbol{\varepsilon}\left(\omega,\mathbf{k}=0\right)\right)=0$, as right after a pole the corresponding eigenvalue becomes negative.

The above considerations, together with Eq. (6), demonstrate that the axion invariant can be solely expressed in terms of the number of poles and positive inertia of the permittivity matrix of the system evaluated at the origin and near infinity:

$$2\frac{\theta}{\pi}=\left(N_{\mathrm{p}}-N_{\mathrm{p}}^{\infty}\right)\Big|_{\omega<\omega_{\mathrm{g}}}+\mathrm{pos}\left(\omega_{\mathrm{g}}\boldsymbol{\varepsilon}\left(\omega_{\mathrm{g}},\mathbf{0}\right)\right)-\mathrm{pos}\left(\omega_{\mathrm{g}}\boldsymbol{\varepsilon}_{\infty}\left(\omega_{\mathrm{g}}\right)\right)\quad(\mathrm{mod}\,4). \tag{10}$$

In the above, it is implicit that $N_{\mathrm{p}}^{\infty}$ refers to the number of poles of $\det\left(\omega\boldsymbol{\varepsilon}\left(\omega,\mathbf{k}\right)\right)$ with $\mathbf{k}$ taken very large but finite. In fact, in the $k\to\infty$ limit some of the material degrees of freedom can become decoupled from the Maxwell equations [7], leading effectively to pole-zero cancellations. Taking $\mathbf{k}$ very large but finite avoids these cancellations and ensures that the pole and zero counting obtained using the strategy delineated above is consistent with the count obtained from the augmented problem operator.

*C. Material responses with a spatial cutoff*

In what follows, I focus on (inversion-symmetric) material responses with the following structure:

$$\mathbf{M}(\omega,\mathbf{k}) = \mathbf{M}^{\text{rec}}(\omega) + \frac{1}{1+k^2/k_{\text{max}}^2}\boldsymbol{\chi}^{\text{cut}}(\omega,\mathbf{k}). \tag{11}$$

Here, $\mathbf{M}^{\text{rec}}(\omega)$ represents a reciprocal inversion-symmetric local response, and $\boldsymbol{\chi}^{\text{cut}}(\omega,\mathbf{k})$ is a generalized susceptibility function that is assumed a polynomial of degree at most one in $\mathbf{k}$. The term $\frac{1}{1+k^2/k_{\text{max}}^2}$ represents a spatial cutoff factor, with $k_{\text{max}}$ the cutoff wave number. Systems with this type of response have been previously studied in the context of Chern insulators in the continuum [7-12]. In particular, when $\mathbf{M}^{\text{rec}}(\omega)$ is taken as the vacuum response, $\mathbf{M}_{\text{vac}} = \begin{pmatrix} \varepsilon_0 \mathbf{1} & 0 \\ 0 & \mu_0 \mathbf{1} \end{pmatrix}$, the full spatial cutoff guarantees that in the short-wavelength limit the material response approaches the vacuum response. For $k \to \infty$, the material degrees of freedom become decoupled from the electromagnetic field, thereby implementing the continuum analogue of an atomic limit. Such a cutoff is consistent with the Kramers-Kronig relations for spatially dispersive media and may also arise naturally from the spatial averaging inherent to effective-medium descriptions [7, 8, 17].

Here, however, a reciprocal part of the response will be left uncut (the term $\mathbf{M}^{\text{rec}}(\omega)$), as otherwise it would not be possible to preserve a full bandgap in the class of considered systems. In fact, for $k$ sufficiently large the material response becomes strongly dominated by the reciprocal piece $\mathbf{M}^{\text{rec}}(\omega)$, which therefore can be identified with $\mathbf{M}_\infty(\omega)$.

Now, the interesting observation is that for the considered system class, the material pole frequencies are independent of *k*, for any finite *k*. This means that, for the considered class, the axion invariant in Eq. (10) can be solely expressed in terms of the positive inertia of the system permittivity matrix:

$$2\frac{\theta}{\pi} = \text{pos}\left(\omega_{\text{g}}\boldsymbol{\varepsilon}\left(\omega_{\text{g}},\mathbf{0}\right)\right) - \text{pos}\left(\omega_{\text{g}}\boldsymbol{\varepsilon}_{\infty}\left(\omega_{\text{g}}\right)\right) \quad (\text{mod}\, 4)\,. \tag{12}$$

Furthermore, for positive $\omega_{\text{g}}$, the factor $\omega_{\text{g}}$ can be dropped from the positive inertia, resulting in:

$$2\frac{\theta}{\pi} = \text{pos}\left(\boldsymbol{\varepsilon}\left(\omega_{\text{g}},\mathbf{0}\right)\right) - \text{pos}\left(\boldsymbol{\varepsilon}_{\infty}\left(\omega_{\text{g}}\right)\right) \quad (\text{mod}\, 4)\,. \tag{13}$$

Note that this final result is restricted to material responses whose poles are independent of the wave vector, as in Eq. (11). Furthermore, due to the reality symmetry of the electromagnetic field, positive- and negative-frequency gaps must have the same axion indicator. Thus, Eq. (13) also applies to negative $\omega_{\text{g}}$.

Appendix A characterizes the band gaps of reciprocal inversion-symmetric materials, with a response $\mathbf{M}(\omega,k)$ that is allowed to depend on $k = |\mathbf{k}|$. It is proven that such systems can have a full gap containing $\omega_{\text{g}}$ shared by the corresponding local asymptotic reference $\mathbf{M}_{\infty}(\omega)$, only when the permittivity at $\omega_{\text{g}}$ is positive definite and the permeability is negative definite for all $k$ (mu-negative or MNG-type response), or alternatively, when the permittivity is negative definite and the permeability is positive definite for all $k$ (epsilon-negative or ENG-type response).

The enunciated property implies that these reciprocal systems have a positive inertia that is independent of the wave vector. Hence, they have a trivial axion topology, consistent with the assumptions underlying the construction of the invariant. In particular, the results of Appendix A imply that for a generic material $\mathrm{pos}\left(\boldsymbol{\varepsilon}_{\infty}\left(\omega_{\mathrm{g}}\right)\right)$ is either 0 or 3.

## III. Gapless Reference for an Axion Phase

If a continuum axion phase exists, it should be possible to reduce it to a gapless phase by letting the strength of the relevant coupling parameter approach zero. Here, I introduce a reciprocal template for such a gapless system with the desired signatures of the axion phase.

Assuming that the gap does not close at the endpoints $k=0$ and $k=\infty$, a nontrivial axion signature requires that the positive inertias of the permittivity at these points differ. For definiteness, I consider an ENG-type response at $\mathbf{k}=\infty$ (implemented through the term $\mathbf{M}^{\mathrm{rec}}\left(\omega\right)$ in Eq. (11)), so that $\mathrm{pos}\left(\boldsymbol{\varepsilon}\left(\omega_{\mathrm{g}},\infty\right)\right)=0$. Then, Eq. (13) shows that to have a nontrivial axion indicator it is necessary that $\mathrm{pos}\left(\boldsymbol{\varepsilon}\left(\omega_{\mathrm{g}},0\right)\right)=2$. In other words, the permittivity must become indefinite in the long-wavelength limit. To achieve this, the term affected by the spatial cutoff, $\boldsymbol{\chi}^{\mathrm{cut}}$, is chosen such that it can provide an uniaxial hyperbolic-type permittivity response (with $\varepsilon_{zz}<0$ and $\varepsilon_t \equiv \varepsilon_{xx}=\varepsilon_{yy}>0$) near $k=0$ in the relevant spectral region. Then, to enforce a gap near $k=0$, one needs a corresponding uniaxial hyperbolic-type permeability response such that $\mu_{zz}>0$ and $\mu_t \equiv \mu_{xx}=\mu_{yy}<0$ near $k=0$. Based on these considerations, I consider the following template:

$$\boldsymbol{\varepsilon}\left(\omega,k\right)=\left(1+\frac{\omega_{pe}^2}{\omega_{0e}^2-\omega^2}\right)\mathbf{1}+s\left(k\right)\frac{\omega_{pe2}^2}{\omega_{0e2}^2-\omega^2}\mathbf{1}_t, \qquad \boldsymbol{\mu}\left(\omega,k\right)=\mathbf{1}+s\left(k\right)\frac{\omega_{pm}^2}{\omega_{0m}^2-\omega^2}\mathbf{1}_t. \quad (14)$$

where $s(k)=1/\left(1+k^2/k_{\max}^2\right)$ is the cutoff factor and $\mathbf{1}_t=\hat{\mathbf{x}}\otimes\hat{\mathbf{x}}+\hat{\mathbf{y}}\otimes\hat{\mathbf{y}}$. I adopt units with $\varepsilon_0=1$ and $\mu_0=1$.

Clearly, the asymptotic ENG reference is described by $\boldsymbol{\varepsilon}_\infty(\omega)=\left(1+\dfrac{\omega_{pe}^2}{\omega_{0e}^2-\omega^2}\right)\mathbf{1}$ and $\boldsymbol{\mu}_\infty(\omega)=\mathbf{1}$. Thus, the band gap near $k=\infty$ is determined by:

$$\omega_{0e}<\omega<\sqrt{\omega_{0e}^2+\omega_{pe}^2}, \qquad \text{(gap at } k=\infty\text{)}. \tag{15a}$$

On the other hand, the gap near the origin must be contained in the region where the transverse permeability is negative ($\mu_t<0$):

$$\omega_{0m}<\omega<\sqrt{\omega_{0m}^2+\omega_{pm}^2}, \qquad (\mu_t<0 \text{ at } k=0). \tag{15b}$$

Hence, the intervals in Eqs. (15a) and (15b) must overlap. In addition, the parameters of the second Lorentz oscillator in Eq. (14), $\omega_{0e2},\omega_{pe2}$, must also guarantee that $\varepsilon_t(\omega,0)>0$ in some range that overlaps the intersection of the intervals in Eqs. (15a) and (15b).

For simplicity, it is assumed that the interval in Eq. (15a) is included in the interval in Eq. (15b) and that the pole of the second Lorentz oscillator satisfies $\omega_{0e2}>\sqrt{\omega_{0e}^2+\omega_{pe}^2}$. In these conditions, the upper frequency bound of the relevant spectral range is determined by the condition $\varepsilon_{zz}=0$, whereas the lower bound is determined by the condition $\varepsilon_t(\omega,0)=0$.

The isofrequency surfaces for the considered uniaxial reciprocal template are determined by two independent polarization branches, specifically

$$\frac{k_t^2}{\varepsilon_{zz}}+\frac{k_z^2}{\varepsilon_t}=\omega^2\mu_t, \qquad \text{and} \qquad \frac{k_t^2}{\mu_{zz}}+\frac{k_z^2}{\mu_t}=\omega^2\varepsilon_t, \tag{16}$$

for transverse magnetic (TM) polarized waves (with magnetic field in the transverse plane) and for transverse electric (TE) polarized waves (with electric field in the transverse plane), respectively. From the theory of Appendix A, the considered template must be gapless in the relevant spectral range, determined by the intersection of the frequency gaps at the two endpoints ($k=0$ and $k=\infty$).

Furthermore, for any frequency in that spectral range, it is possible to construct longitudinal-type solutions with $k_z=0$ of electric and magnetic character, respectively, using the same strategy as in Appendix A. These solutions are found by solving $\varepsilon_t(\omega,k_t)=0$ ($\mu_t(\omega,k_t)=0$) for the TM-polarization (TE-polarization) case. Note that by construction these two functions have opposite signs at the two endpoints $k=0$ and $k=\infty$.

There is always some critical frequency, $\omega_c$, in the relevant spectral range such that $\varepsilon_t(\omega_c,k_t)=0=\mu_t(\omega_c,k_t)$ for some common $k_t$. To show this, note that $\varepsilon_t(\omega,k)=\varepsilon_{zz}(\omega)+s(k)\chi_e^{\text{cut}}(\omega)$ and $\mu_t(\omega,k)=\mu_{zz}(\omega)+s(k)\chi_m^{\text{cut}}(\omega)$, with $\chi_e^{\text{cut}}(\omega)>0$ and $\chi_m^{\text{cut}}(\omega)<0$. The cutoff factors for the electric and magnetic longitudinal modes are then $s_e=\dfrac{-\varepsilon_{zz}(\omega)}{\chi_e^{\text{cut}}(\omega)}$ and $s_m=\dfrac{-\mu_{zz}(\omega)}{\chi_m^{\text{cut}}(\omega)}$, respectively. As already noted, the upper bound of the relevant spectral range is determined by the condition $\varepsilon_{zz}(\omega)=0$, whereas the lower bound is determined by the condition $\varepsilon_{zz}(\omega)+\chi_e^{\text{cut}}(\omega)=0$. Thus, $s_e$ varies over the full range $[0,1]$ in the relevant interval. This demonstrates that the two cutoff factors $s_e$ and $s_m$ must indeed intersect for some frequency $\omega_c$. In particular, the identity $\dfrac{\varepsilon_{zz}(\omega_c)}{\mu_{zz}(\omega_c)}=\dfrac{\chi_e^{\text{cut}}(\omega_c)}{\chi_m^{\text{cut}}(\omega_c)}$ holds true.

Remarkably, at the critical frequency $\omega_c$ the dispersion of the two polarization branches becomes exactly degenerate and the isofrequency surfaces for TE and TM waves coincide. To demonstrate this, I note that at $\omega_c$ the transverse permittivity can be written as $\varepsilon_t(\omega_c,k)=\varepsilon_{zz}(\omega_c)+s(k)\frac{\varepsilon_{zz}(\omega_c)}{\mu_{zz}(\omega_c)}\chi_m^{\text{cut}}(\omega_c)$ so that $\varepsilon_t(\omega_c,k)=\frac{\varepsilon_{zz}(\omega_c)}{\mu_{zz}(\omega_c)}\mu_t(\omega_c,k)$. Substituting this relation into the TE-dispersion equation in Eq. (16), one readily finds that it coincides with the TM-dispersion at $\omega=\omega_c$ for any wave vector.

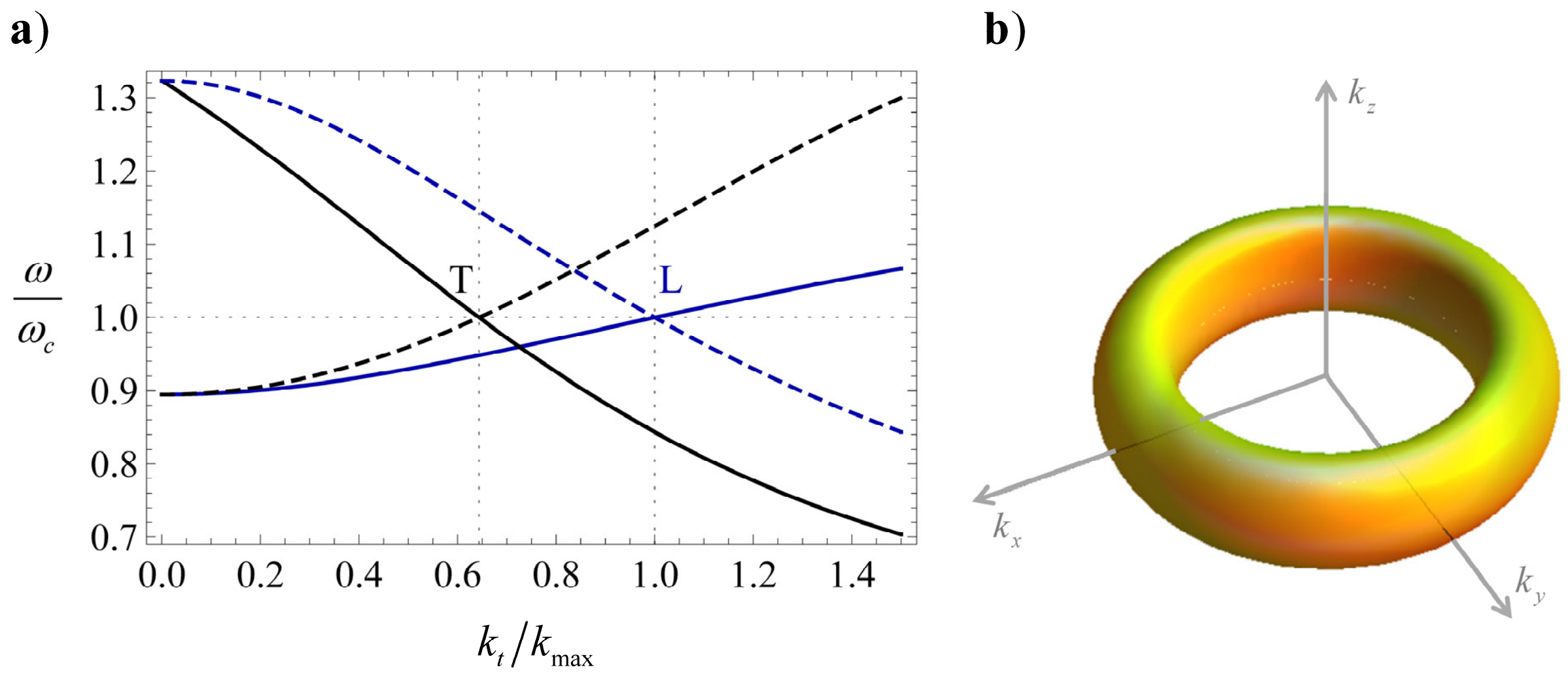

**Fig.** 1 (a) Dispersion diagram $\omega$ versus $k_t$ for TM-polarized waves (solid lines) and TE-polarized waves (dashed lines). The longitudinal (transverse) branches are plotted with blue (black) color. The material parameters are $\omega_{pe}=\omega_{pm}=\sqrt{6}\omega_{0e}$, $\omega_{0m}=\omega_{0e}$, $\omega_{0e2}=3\omega_{0e}$, $\omega_{pe2}=\sqrt{10}\omega_{0e}$ and $k_{\max}=2\omega_{0e}/c$, corresponding to the critical frequency $\omega_c=2\omega_{0e}$ marked with the horizontal gray line. The L point denotes the crossing of the two longitudinal branches, whereas the T point denotes the crossing of the two transverse branches. The wave numbers $k_t$ associated with these crossings are marked with the vertical gray lines. (b) Toroidal-type isofrequency surface in momentum space at the critical frequency. The TM and TE isofrequency surfaces are exactly coincident.

To illustrate the discussion, Fig. 1a depicts the dispersion of the TE and TM polarized waves for $k_z = 0$. The gray horizontal line marks the critical frequency. As seen, the TE and TM branches cross at two different values of $k_t$ at the critical frequency. The point L corresponds to the longitudinal branches crossing (blue lines) discussed earlier. Conversely, the point T corresponds to the crossing of the two transverse branches (black lines), determined by the solutions of $k_t^2 = \omega^2 \mu_t \varepsilon_{zz}$ and $k_t^2 = \omega^2 \varepsilon_t \mu_{zz}$ for TM and TE waves, respectively.

It is underlined that the coincidence of the two crossings at the same frequency is not a result of fine-tuning the material parameters, but rather a general property of the proposed system: the dispersion of TE and TM branches always collapses into a single isofrequency surface at the critical frequency. This degenerate isofrequency surface is represented in Fig. 1b in momentum space. As seen, it has a continuous rotational symmetry about the *z*-axis and has the shape of a torus. Related toroidal isofrequency surfaces and nodal degeneracies have been studied in periodic photonic systems [41-43].

Figure 2 shows sections of the TE and TM isofrequency surfaces in the $k_y = 0$ plane, for different frequencies and the same material parameters as in Fig. 1. As can be seen, slightly below the critical frequency (panel *a*) or slightly above the critical frequency (panel *c*), the two isofrequency surfaces separate from each other. Below the critical frequency the TM isofrequency toroidal surface is contained inside the TE isofrequency toroidal surface, whereas above the critical frequency the opposite behavior is observed. In the central panel of Fig. 2, corresponding to the critical case, the two isofrequency surfaces become coincident.

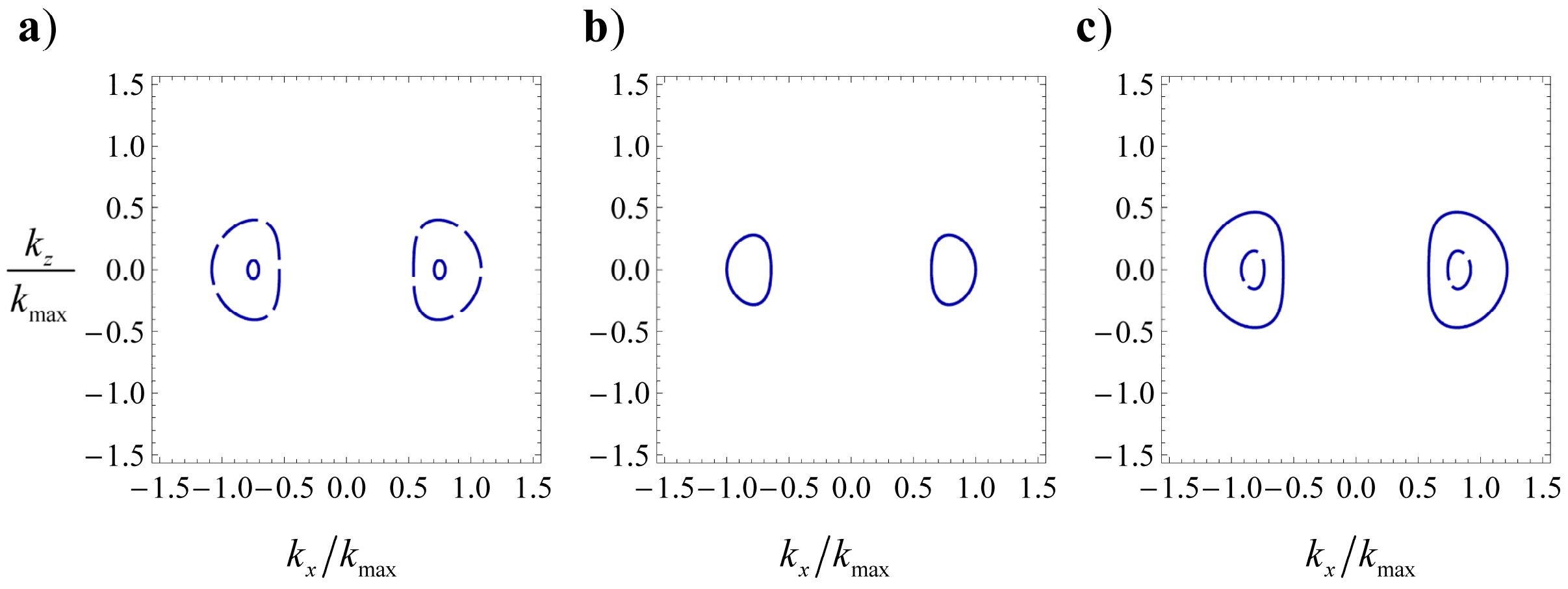

**Fig.** 2 Isofrequency contours in the $k_y = 0$ plane at the frequency **a)** $\omega = 0.97\omega_c$, **b)** $\omega = \omega_c$ and **c)** $\omega = 1.03\omega_c$. The material parameters are the same as in Fig. 1. Solid lines: TM polarization. Dashed lines: TE polarization.

As will be shown in the following sections, an axion phase can be created near $\omega = \omega_c$ by perturbing the considered template response with a suitable nonreciprocal coupling that hybridizes and spectrally separates the two coincident TE and TM isofrequency surfaces. Here, the coincident isofrequency surfaces play the role of a Dirac point in two-dimensional insulators, which under a suitable nonreciprocal perturbation can become gapped and create a Chern phase. Importantly, in the current three-dimensional system the perturbation needs to gap the entire isofrequency surface, rather than just a single point.

## IV. Perturbation Theory

In this section, I investigate under which conditions a material perturbation can gap the reciprocal axion template near the critical frequency. To this end, I first develop a first-order perturbation theory that describes how the spectrum of the material is modified by a nonreciprocal coupling.

### *A. First-order perturbation theory*

As seen in Sec. III, at the critical frequency $\omega = \omega_c$ the spectrum of the reciprocal uniaxial template consists of two coincident toroidal isofrequency surfaces associated with TE- and TM-polarized waves. I denote the material matrix of the reciprocal template as $\mathbf{M}_0(\omega,\mathbf{k})$ and consider a perturbed (Hermitian) system described by

$$\mathbf{M}(\omega,\mathbf{k}) = \mathbf{M}_0(\omega,\mathbf{k}) + \delta\mathbf{M}(\omega,\mathbf{k}), \tag{17}$$

where $\delta\mathbf{M}$ is the perturbation.

Appendix B develops a two-band model to characterize how the perturbation changes the system spectrum near the critical frequency. By expanding the electromagnetic fields in terms of the unperturbed TE and TM eigenstates, denoted by $\mathbf{f}_{1,\mathbf{k}}$ and $\mathbf{f}_{2,\mathbf{k}}$, respectively, ($\mathbf{f} = \alpha_1\mathbf{f}_{1,\mathbf{k}} + \alpha_2\mathbf{f}_{2,\mathbf{k}}$), one arrives at the eigensystem

$$\begin{pmatrix} \mathbf{f}^*_{1,\mathbf{k}} \cdot \delta\mathbf{M} \cdot \mathbf{f}_{1,\mathbf{k}} & \mathbf{f}^*_{1,\mathbf{k}} \cdot \delta\mathbf{M} \cdot \mathbf{f}_{2,\mathbf{k}} \\ \mathbf{f}^*_{2,\mathbf{k}} \cdot \delta\mathbf{M} \cdot \mathbf{f}_{1,\mathbf{k}} & \mathbf{f}^*_{2,\mathbf{k}} \cdot \delta\mathbf{M} \cdot \mathbf{f}_{2,\mathbf{k}} \end{pmatrix} \begin{pmatrix} \alpha_1 \\ \alpha_2 \end{pmatrix} = \begin{pmatrix} -\dfrac{\omega - \omega_{1,\mathbf{k}}}{\omega_c} & 0 \\ 0 & -\dfrac{\omega - \omega_{2,\mathbf{k}}}{\omega_c} \end{pmatrix} \begin{pmatrix} \alpha_1 \\ \alpha_2 \end{pmatrix}. \tag{18}$$

This approach provides a description of the spectrum within first-order perturbation theory. The material perturbation is evaluated at the critical frequency: $\delta\mathbf{M} \equiv \delta\mathbf{M}(\omega_c,\mathbf{k})$. The unperturbed eigenstates are normalized as in Eq. (B4).

The secular equation associated with the above Hermitian eigensystem is

$$\left(\delta m_{\mathrm{s},\mathbf{k}} + \frac{\omega - \omega_{\mathrm{s},\mathbf{k}}}{\omega_c}\right)^2 - \left(\delta m_{\mathrm{d},\mathbf{k}} - \frac{\omega_{\mathrm{d},\mathbf{k}}}{\omega_c}\right)^2 - \left|\delta m_{12}\right|^2 = 0, \tag{19}$$

where I introduce

$$\delta m_{\mathrm{s},\mathbf{k}} = \frac{\delta m_{11} + \delta m_{22}}{2}, \qquad \delta m_{\mathrm{d},\mathbf{k}} = \frac{\delta m_{11} - \delta m_{22}}{2}, \tag{20a}$$

$$\omega_{\mathrm{s},\mathbf{k}} = \frac{\omega_{1,\mathbf{k}} + \omega_{2,\mathbf{k}}}{2}, \qquad \omega_{\mathrm{d},\mathbf{k}} = \frac{\omega_{1,\mathbf{k}} - \omega_{2,\mathbf{k}}}{2}, \tag{20b}$$

with $\delta m_{ij} \equiv \mathbf{f}_{i,\mathbf{k}}^* \cdot \delta\mathbf{M} \cdot \mathbf{f}_{j,\mathbf{k}}$. Solving the secular equation, one obtains:

$$\frac{\omega - \omega_{\mathrm{s},\mathbf{k}}}{\omega_c} = -\delta m_{\mathrm{s},\mathbf{k}} \pm \sqrt{\left(\delta m_{\mathrm{d},\mathbf{k}} - \frac{\omega_{\mathrm{d},\mathbf{k}}}{\omega_c}\right)^2 + \left|\delta m_{12}\right|^2}\,. \tag{21}$$

When the radicand vanishes somewhere in the vicinity of the degenerate isofrequency surface, the two branches remain degenerate to first order. In that case, a gap may still open at higher order.

Thus, below I focus on the conditions that guarantee a local gap to first order, i.e., ensure that the radicand remains strictly positive in the vicinity of the degenerate isofrequency surface. It should be noted that a direct gap does not always guarantee a global gap because the shift term $\omega_{\mathrm{s},\mathbf{k}} - \omega_c \delta m_{\mathrm{s},\mathbf{k}}$ is wave-vector dependent: an indirect gap overlap remains possible. Hence, even if the radicand remains strictly positive, one still needs to verify that the gap is global.

It is convenient to write the wave vector in the vicinity of the degenerate isofrequency surface (Fig. 1b) as $\mathbf{k} = \mathbf{k}_\mathrm{S} + q\hat{\mathbf{n}}_\mathrm{S}$, where $\mathbf{k}_\mathrm{S}$ is a vector on the surface and $\hat{\mathbf{n}}_\mathrm{S}$ is normal to the surface. Furthermore, in the vicinity of the degenerate isofrequency surface

$$\omega_{\mathrm{d},\mathbf{k}} = \frac{\omega_{1,\mathbf{k}} - \omega_{2,\mathbf{k}}}{2} \approx q\hat{\mathbf{n}}_\mathrm{S} \cdot \nabla_\mathbf{k} \omega_{\mathrm{d},\mathbf{k}} = -q\left|\nabla_\mathbf{k} \omega_{\mathrm{d},\mathbf{k}}\right|. \tag{22}$$

I take into account that $\nabla_\mathbf{k} \omega_{i,\mathbf{k}}$ is normal to the isofrequency surface and assume that $\hat{\mathbf{n}}_\mathrm{S}$ points outward from the toroidal surface. Note that for the considered reciprocal template $\nabla_\mathbf{k} \omega_{\mathrm{TE},\mathbf{k}}$ ($\nabla_\mathbf{k} \omega_{\mathrm{TM},\mathbf{k}}$) is anti-parallel (parallel) to $\hat{\mathbf{n}}_\mathrm{S}$ (see the isofrequency contours in Fig. 2), and therefore $\left|\nabla_\mathbf{k} \omega_{\mathrm{d},\mathbf{k}}\right|$ is nonzero. The above formula shows that a small displacement in the direction perpendicular to the isofrequency surface (with $q = -\omega_c \delta m_{\mathrm{d},\mathbf{k}} / \left|\nabla_{\mathbf{k}} \omega_{\mathrm{d},\mathbf{k}}\right|$) can always nullify the first term of the radicand in Eq. (21). Hence, it follows that the system can have a local (direct) gap to first order only if $\delta m_{12}$ is strictly nonzero for all $\mathbf{k}_{\mathrm{S}}$.

In summary, the preceding analysis demonstrates that the system becomes locally gapped within first order perturbation theory if and only if:

$$\mathbf{f}_{1,\mathbf{k}}^{*} \cdot \delta\mathbf{M} \cdot \mathbf{f}_{2,\mathbf{k}} \neq 0, \text{ for all } \mathbf{k} \text{ on the toroidal surface.} \quad (23)$$

Again, it is emphasized that the above condition does not guarantee a global gap because the center frequency ($\omega_{\mathrm{s},\mathbf{k}} - \omega_c \delta m_{\mathrm{s},\mathbf{k}}$) may in general vary along the toroidal surface.

*B. Perturbations of the permittivity and permeability responses*

Next, I apply the theory of the previous subsection to perturbations of the permittivity and permeability responses of the reciprocal template such that $\delta\mathbf{M} = \begin{pmatrix} \delta\boldsymbol{\varepsilon} & 0 \\ 0 & \delta\boldsymbol{\mu} \end{pmatrix}$. Here, the permittivity and permeability perturbations $\delta\boldsymbol{\varepsilon}$ and $\delta\boldsymbol{\mu}$ are arbitrary Hermitian matrices that may depend on the wave vector and frequency, such that the total system response remains in the class defined by Eq. (11). Note also that they must remain globally bounded throughout wave-vector space, as any singularities for real $\mathbf{k}$ would violate the assumed regularity of the material response.

Unfortunately, independent of the specific structure of the considered perturbation, the direct gap criterion in Eq. (23) cannot be satisfied globally on the toroidal surface. This is a straightforward consequence of the fact that the proposed reciprocal template is associated with longitudinal solutions on the outer circle in the $k_z = 0$ plane (corresponding to point L in Fig.

1a). These solutions were already discussed in Sec. III, and occur when $\varepsilon_t(\omega,k_t)=0$ for TM-polarized waves, and $\mu_t(\omega,k_t)=0$ for TE-polarized waves. For such longitudinal solutions, the electromagnetic modes $\mathbf{f}_{\mathbf{k}}=\begin{pmatrix}\mathbf{E}_{\mathbf{k}} & \mathbf{H}_{\mathbf{k}}\end{pmatrix}^T$ have the following structure: $\mathbf{f}_{1,\mathbf{k}}=\mathbf{f}_{\mathrm{TE},\mathbf{k}}\sim\begin{pmatrix}0 & \hat{\mathbf{k}}\end{pmatrix}^T$ and $\mathbf{f}_{2,\mathbf{k}}=\mathbf{f}_{\mathrm{TM},\mathbf{k}}\sim\begin{pmatrix}\hat{\mathbf{k}} & 0\end{pmatrix}^T$, with $\hat{\mathbf{k}}$ the unit vector, which for the longitudinal modes is in the $k_z=0$ plane. Clearly, due to the diagonal structure of $\delta\mathbf{M}$, one has $\mathbf{f}_{1,\mathbf{k}}^*\cdot\delta\mathbf{M}\cdot\mathbf{f}_{2,\mathbf{k}}=0$ on the longitudinal mode circle of the toroidal surface. Thus, within the validity of first-order perturbation theory, the system always remains gapless, independent of the structure of the perturbations $\delta\boldsymbol{\varepsilon}$ and $\delta\boldsymbol{\mu}$.

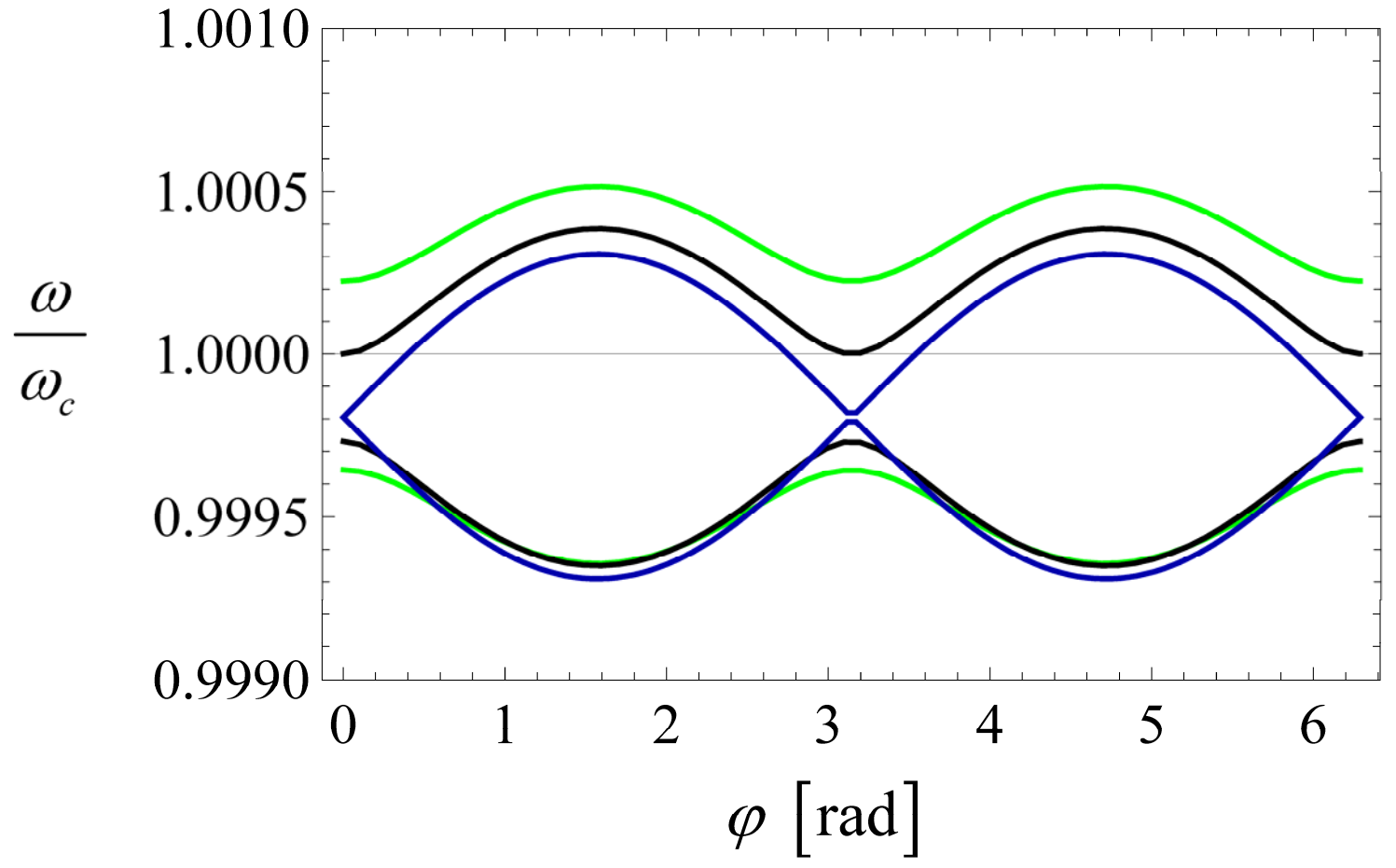

**Fig.** 3 Dispersion diagram $\omega$ versus $\varphi$ for a gyrotropic perturbation of the reciprocal template and $\mathbf{k}=(k_t\cos\varphi,k_t\sin\varphi,0)$ with $k_t\approx k_L$. Here, $k_L$ is the radius of the longitudinal-mode circle of the reciprocal template. Green lines: $k_t=0.9994k_L$; Black lines: $k_t=k_L$; Blue lines: $k_t=1.00052k_L$. The horizontal gray line marks the critical frequency. The reciprocal template has the same parameters as in Fig. 1. The perturbation is such that $\delta\boldsymbol{\varepsilon}=i\delta_\varepsilon s_k\hat{\mathbf{n}}_\varepsilon\times\mathbf{1}$ and $\delta\boldsymbol{\mu}=i\delta_\mu s_k\hat{\mathbf{n}}_\mu\times\mathbf{1}$ with $\delta_\varepsilon=\delta_\mu=0.1$ and $\hat{\mathbf{n}}_\varepsilon$ and $\hat{\mathbf{n}}_\mu$ unit vectors defined in polar coordinates by the angles $\varphi_\varepsilon=45^{\circ}$, $\theta_\varepsilon=30^{\circ}$ and $\varphi_\mu=0^{\circ}$, $\theta_\mu=90^{\circ}$, respectively.

Numerical simulations support the conclusions of first-order perturbation theory, as they suggest that it is not possible to gap the reciprocal template using purely electric and purely magnetic perturbations. Note that for the system class defined by Eq. (11), the nonreciprocal parts of $\delta\boldsymbol{\varepsilon}$ and $\delta\boldsymbol{\mu}$ necessarily have a gyrotropic structure, analogous to the response provided by magnetically biased electric or magnetic materials.

As an illustration of the preceding discussion, Fig. 3 shows the band diagram of the reciprocal template with gyrotropic nonreciprocal perturbations of the type $\delta\boldsymbol{\varepsilon} = i\delta_\varepsilon s_k \hat{\mathbf{n}}_\varepsilon \times \mathbf{1}$ and $\delta\boldsymbol{\mu} = i\delta_\mu s_k \hat{\mathbf{n}}_\mu \times \mathbf{1}$. Here, $\hat{\mathbf{n}}_\varepsilon$ and $\hat{\mathbf{n}}_\mu$ determine the symmetry axis of the electric and magnetic gyrotropic perturbations, respectively, and $s_k = 1/\left(1 + k^2 / k_{\max}^2\right)$ is the cutoff factor. For simplicity, the frequency dispersion of the perturbations is neglected here. Figure 3 shows the spectrum in the vicinity of the outer circle of the unperturbed toroidal surface, associated with the unperturbed longitudinal modes. As seen, even though the perturbation lifts the degeneracy for most directions in the $k_z = 0$ plane (due to higher-order corrections), the gap remains closed near $\varphi = 0^\circ$ and $\varphi = 180^\circ$, corresponding to the $k_x$ axis. I checked that a similar trend is maintained for other orientations of $\hat{\mathbf{n}}_\varepsilon$ and $\hat{\mathbf{n}}_\mu$, which suggests that the proposed reciprocal template cannot be gapped using gyrotropic perturbations.

Furthermore, it is demonstrated in Appendix C that for arbitrary electric and magnetic gyrotropic systems with a continuous rotational symmetry about the *z*-axis (corresponding to gyrotropic responses described by $\hat{\mathbf{n}}_\varepsilon = \hat{\mathbf{n}}_\mu = \hat{\mathbf{z}}$), it is impossible to create an axion phase. The result of Appendix C is general, accounts for material dispersion, and does not rely on perturbation theory.

## V. Odd Magnetoelectric Coupling

The negative results of the previous section are somewhat surprising, as they suggest that gyrotropic electric and/or magnetic responses may not provide sufficient degrees of freedom to realize an axion phase in the continuum. This is especially noteworthy because previous proposals of photonic axion insulators rely precisely on gyrotropic materials [34-36]. A common feature of these proposals is the use of compensated gyrotropic arrangements, with nonreciprocal elements having oppositely oriented gyrotropy axes within the unit cell. These arrangements are natural because a three-dimensional axion insulator requires the vanishing of the Chern numbers, and hence the suppression of the net Hall-type response. In an effective-medium description, the compensation indicates a weak or vanishing conventional gyrotropic response. However, since the photonic axion phases require broken time-reversal symmetry, this suggests that nonreciprocal mechanisms beyond conventional gyrotropy may play an important role in a continuum theory.

To understand what types of nonreciprocal responses may be relevant, it is useful to note the analogy between photonic crystals with compensated gyrotropy and natural magnetic materials with compensated magnetic order, such as antiferromagnets and altermagnets. Related analogies have recently been explored in the context of photonic altermagnetism [45, 46]. Here, the analogy is used to identify the electromagnetic responses that may arise in an effective-medium description of crystalline photonic axion insulators.

In conventional collinear antiferromagnets, the oppositely oriented magnetic sublattices may be related by a lattice translation combined with time reversal. Such a symmetry enforces reciprocity of the effective bulk electromagnetic response, and thus cannot be the basis of the nonreciprocal continuum axion phase considered here. Another possibility is the preservation of the combined inversion and time-reversal symmetry, as in $Cr_2O_3$, which permits a nonreciprocal (Tellegen-type) magnetoelectric response [47-48]. However, this requires inversion symmetry to be broken, making such systems unsuitable for the purpose of this work.

A different situation arises in altermagnets [49-52], where the oppositely oriented magnetic moments are related by crystal rotations rather than translations or inversion. Centrosymmetric altermagnets are particularly relevant in this context as they preserve inversion while breaking time-reversal symmetry. Their optical response may be nonreciprocal, and conventional gyrotropy is allowed in certain symmetry classes. Importantly, their nonreciprocal response need not be restricted to conventional gyrotropy [51-52]. Specifically, nonreciprocal magnetoelectric responses involving spatial gradients of the electromagnetic fields have been discussed in these systems [52].

This observation is especially relevant from an effective-medium perspective. Indeed, the optical response of centrosymmetric altermagnets should be qualitatively similar to the effective response of inversion-symmetric photonic crystals with compensated gyrotropy. In both cases, inversion symmetry is preserved while time-reversal symmetry is broken, despite the compensation of the magnetic or gyrotropic contributions. In an inversion-symmetric medium, the permittivity and permeability tensors must be even functions of the wave vector, whereas the magnetoelectric tensors must be odd. Consequently, assuming an analytic response, the leading spatial-dispersion corrections to the electric and magnetic responses are quadratic in the wave vector, while magnetoelectric coupling is allowed already at first order [53-54]. Hence, besides any residual local gyrotropy, odd magnetoelectric coupling is the leading spatially dispersive correction allowed by inversion symmetry [52].

Motivated by the analogy with centrosymmetric altermagnets, and considering that conventional gyrotropic perturbations were insufficient to open a full gap in the proposed reciprocal template, I next investigate whether a nonreciprocal odd magnetoelectric response, analogous to that naturally present in some centrosymmetric altermagnets [52], can support a nontrivial continuum axion phase.

### *A. Magnetoelectric response*

I consider an electromagnetic continuum described by a material matrix with the structure in Eq. (3) in the general class of systems with a spatial cutoff [Eq. (11)]. As already discussed in Sec. II.B, inversion symmetry imposes the constraint $\mathbf{M}(\omega,\mathbf{k}) = \boldsymbol{\sigma}_z \cdot \mathbf{M}(\omega,-\mathbf{k}) \cdot \boldsymbol{\sigma}_z$, which forbids a local magnetoelectric response. However, inversion symmetry remains compatible with an odd magnetoelectric coupling such that $\boldsymbol{\xi}(\omega,\mathbf{k}) = -\boldsymbol{\xi}(\omega,-\mathbf{k})$ and $\boldsymbol{\zeta}(\omega,\mathbf{k}) = -\boldsymbol{\zeta}(\omega,-\mathbf{k})$ [53]. To explore this type of coupling, consider a general structure of the form

$$\boldsymbol{\xi}(\omega,\mathbf{k}) = \frac{s(k)}{k_{\max}}\left(\boldsymbol{\xi}_x k_x + \boldsymbol{\xi}_y k_y + \boldsymbol{\xi}_z k_z\right), \qquad \boldsymbol{\zeta}(\omega,\mathbf{k}) = \frac{s(k)}{k_{\max}}\left(\boldsymbol{\zeta}_x k_x + \boldsymbol{\zeta}_y k_y + \boldsymbol{\zeta}_z k_z\right), \quad (24)$$

where the matrices $\boldsymbol{\zeta}_i$ and $\boldsymbol{\xi}_i$ are linked by $\boldsymbol{\xi}_i = (\boldsymbol{\zeta}_i)^\dagger$ to ensure that the system response remains Hermitian ($\boldsymbol{\xi}(\omega,\mathbf{k}) = \left[\boldsymbol{\zeta}(\omega,\mathbf{k})\right]^\dagger$). The cutoff factor $s(k)$ ensures that the (potentially nonreciprocal) magnetoelectric coupling tensors do not contribute to the asymptotic response $\mathbf{M}_\infty(\omega)$. Recall that the adopted units are such that $\varepsilon_0 = 1$ and $\mu_0 = 1$, so the material matrix is dimensionless.

In order that the magnetoelectric coupling breaks time-reversal symmetry, it is necessary that $\xi(\omega,\mathbf{k}) \neq -[\zeta(\omega,-\mathbf{k})]^T = [\zeta(\omega,\mathbf{k})]^T$ [37]. Using $\xi(\omega,\mathbf{k}) = [\zeta(\omega,\mathbf{k})]^\dagger$, one sees that a nonreciprocal response requires:

$$\xi(\omega,\mathbf{k}) \neq [\xi(\omega,\mathbf{k})]^* \qquad \text{(nonreciprocal response).} \tag{25}$$

Thus, nontrivial phases can be feasible only when $\xi(\omega,\mathbf{k})$ is complex-valued.

In Appendix D, I specialize the first-order gapping condition [Eq. (23)] to the case of odd-magnetoelectric coupling perturbations of the reciprocal template. It is demonstrated that, unlike purely electric or purely magnetic perturbations, failure of the first-order gapping criterion for a nonreciprocal magnetoelectric perturbation is much more demanding and generally requires simultaneous zeros of two independent real functions. This suggests that odd magnetoelectric perturbations can be the basis to engineer a photonic axion phase in an electromagnetic continuum.

*B. Numerical example*

In order to illustrate how the odd magnetoelectric coupling can unlock the axion phase, consider a material response of the form:

$$\xi(\omega,\mathbf{k}) = \xi_0 \frac{s(k)}{k_{\max}} \begin{pmatrix} k_x + ik_y & 0 & 0 \\ 0 & k_x - ik_y & 0 \\ 0 & 0 & k_z \end{pmatrix}, \qquad \text{for } \omega \approx \omega_c. \tag{26}$$

For simplicity, frequency dispersion in the magnetoelectric response is neglected so that Eq. (26) applies only in the vicinity of the critical frequency. The permittivity and permeability responses will be taken as those of the gapless reciprocal template, which has the required inversion-parity signature for a nontrivial axion phase once a full gap is opened.

The considered magnetoelectric coupling, when combined with the vacuum response, is compatible with a positive material energy ($\partial_\omega[\omega\mathbf{M}]$ is positive definite [7]) provided the dimensionless parameter $\xi_0$ satisfies $|\xi_0| < 2$. The tensor $\boldsymbol{\xi}(\omega,\mathbf{k})$ describes a material with no additional spatial symmetries (no mirrors, no rotational symmetries), besides the inversion symmetry, corresponding to the point group $\bar{1}$.

I numerically checked that a magnetoelectric response with the structure of Eq. (26) satisfies the gap condition [Eq. (D4) in Appendix D], and thus first-order perturbation theory guarantees that the TM- and TE- hybridized bands do not touch in the vicinity of the original toroidal surface. Furthermore, I have computed the exact band structure of the material and numerically confirmed that the odd-magnetoelectric response indeed opens a global gap. This is shown in Fig. 4, which represents the projected band structure of the material for $\xi_0 = 0$ (top panel, corresponding to the same reciprocal template as in Fig. 1) and for $\xi_0 = 0.2$ (bottom panel). The projected band structure represents $\omega$ versus the in-plane wave vector magnitude $k_t$. The shaded regions correspond to the projection of the function $\omega(k_t;k_z,\varphi)$ for all possible values of $k_z$ and $\varphi$, where $\varphi$ determines the orientation of the in-plane wave vector. As seen, in the absence of the odd-magnetoelectric coupling, the toroidal isofrequency surface at the critical frequency is mapped into a line segment delimited by the vertical gray lines. With the addition of the nonreciprocal coupling, the two shaded regions separate from each other, creating a topological gap near the critical frequency $\omega_c$, defined by $0.992\omega_c < \omega < 1.008\omega_c$. As the perturbation does not affect the positive inertia of the permittivity matrix at the endpoints ($k = 0$ and $k = \infty$), the inversion-parity indicator is the same as for the reciprocal template, corresponding to a nontrivial axion phase of the fully gapped system: $\theta = \pi$. Thereby, the considered system realizes the axion phase in an electromagnetic continuum.

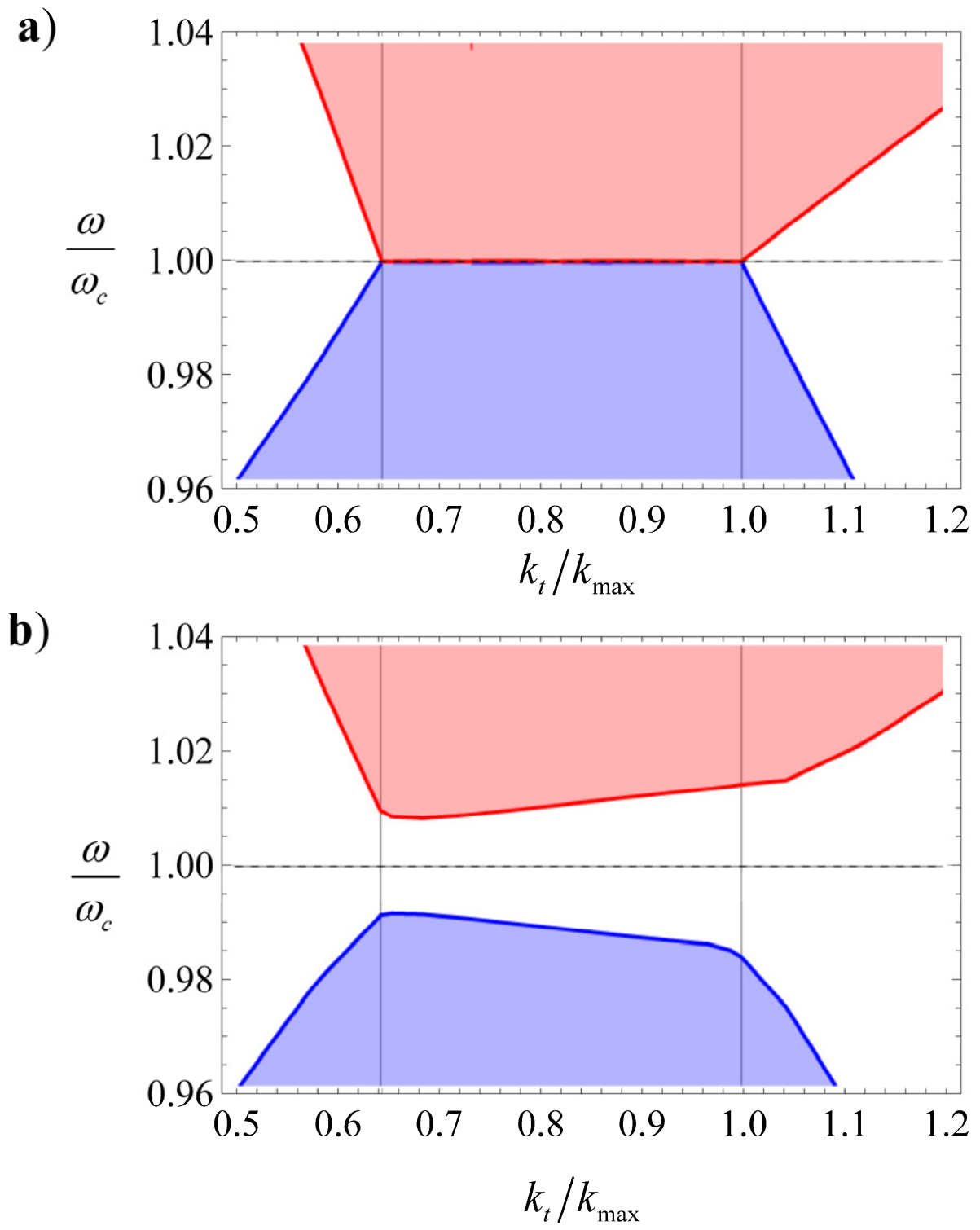

**Fig.** 4 Projected diagram $\omega = \omega(k_t; k_z, \varphi)$ as a function of $k_t$. Top panel: reciprocal template of Fig. 1. Bottom panel: reciprocal template perturbed by an odd magnetoelectric coupling as in Eq. (26) with $\xi_0 = 0.2$. The gray vertical gridlines mark the inner and outer radii of the toroidal isofrequency surface of the reciprocal template in the $k_z = 0$ plane. The horizontal gridline marks the critical frequency.

Additional numerical simulations show that the same behavior is observed for other nonreciprocal odd-magnetoelectric perturbations, even in the presence of additional symmetries. For example, perturbations compatible with the $\bar{3}m'$ magnetic point group can also open nontrivial axion phases (not shown here). These simulations suggest that the lower the symmetry, the easier it is to gap the reciprocal template.

# VI. Conclusion

In summary, I have extended the concept of crystalline photonic axion insulators to electromagnetic continua. In particular, I have introduced a material regularization based on a wave-vector cutoff, such that the material response becomes reciprocal at large wave vectors, and derived an inversion-symmetry indicator for the axion phase. The symmetry indicator depends only on the positive inertia (i.e., the number of positive eigenvalues) of the permittivity matrix at the origin and at infinity in wave-vector space.

Then, I introduced a gapless reciprocal template with an inversion-parity signature compatible with an axion phase, and explored different types of nonreciprocal couplings to open a complete gap and create an axion phase. In particular, a first-order perturbation theory was developed to study the effect of a generic inversion-symmetric perturbation on the reciprocal template. The analysis shows that, to first order, purely electric or magnetic perturbations cannot gap the reciprocal template. Numerical simulations suggest that this obstruction persists beyond first order for gyrotropic perturbations. Furthermore, it was rigorously demonstrated that rotationally symmetric gyrotropic systems within the considered material class cannot support a nontrivial axion phase.

To understand which mechanisms may be appropriate to engineer an axion phase, I discussed the effective-medium analogues of known crystalline photonic axion insulators. The discussion establishes an interesting connection between the effective response of these systems and the optical response of centrosymmetric altermagnets. In such magnetic materials, compensation of the magnetic order can suppress conventional gyrotropy, while nonreciprocal effects may persist through spatially dispersive odd-magnetoelectric responses involving gradients of the fields.

Inspired by this connection, I studied odd magnetoelectric perturbations of the reciprocal template and demonstrated that they can open a complete frequency gap with a nontrivial axion invariant, providing a new paradigm for engineering axion phases in photonic systems.

Looking forward, it will be interesting to investigate whether the proposed continua can support chiral hinge states and to develop an appropriate theoretical framework for describing them. In addition, the present study suggests that, just as crystalline photonic Chern insulators have continuum counterparts in gyrotropic electromagnetic media, crystalline photonic axion insulators may have natural continuum counterparts in centrosymmetric altermagnets. Thus, naturally occurring altermagnets could, in principle, support a photonic axion phase in the continuum. How far this analogy can be taken, and whether any naturally occurring altermagnet can be regarded as a continuum photonic axion insulator, are relevant open questions.

**Acknowledgements:** This work is supported in part by the Institution of Engineering and Technology (IET), by the Simons Foundation (award SFI-MPS-EWP-00008530-10), and by national funds through FCT – Fundação para a Ciência e a Tecnologia, I.P., and, when eligible, co-funded by EU funds under project/support UID/50008/2025 – Instituto de Telecomunicações, with DOI identifier https://doi.org/10.54499/UID/50008/2025.

## Appendix A: Properties of Gapped Reciprocal Systems

In this Appendix, I characterize gapped lossless reciprocal systems described by a material matrix of the form

$$\mathbf{M}(\omega,\mathbf{k})=\begin{pmatrix}\boldsymbol{\varepsilon}(\omega,k) & 0\\ 0 & \boldsymbol{\mu}(\omega,k)\end{pmatrix}. \tag{A1}$$

The dependence of the tensors on the wave vector is restricted to the magnitude $k=|\mathbf{k}|$. Reciprocity implies that the permittivity and permeability tensors are symmetric under transposition. Since the system is also assumed to be conservative, the material matrix is Hermitian. Together, these properties imply that the two tensors are real symmetric.

To begin with, I show that $\det\boldsymbol{\varepsilon}(\omega,k)$ and $\det\boldsymbol{\mu}(\omega,k)$ cannot vanish in the gap for any *k*. In fact, if they did, for example if $\det\boldsymbol{\varepsilon}(\omega_{\mathrm{g}},k_c)=0$ for some $k_c$ with $\omega_{\mathrm{g}}$ in the relevant gap, then it would be possible to pick $\mathbf{E}_0$ nontrivial and real-valued such that:

$$\boldsymbol{\varepsilon}(\omega_{\mathrm{g}},k_c)\cdot\mathbf{E}_0=0\,. \tag{A2}$$

In that case, one can construct a solution of the homogeneous Maxwell's equations [Eq. (7)],

$$\begin{aligned}\mathbf{k}\times\mathbf{E}&=\omega\boldsymbol{\mu}(\omega,k)\cdot\mathbf{H}\\ \mathbf{k}\times\mathbf{H}&=-\omega\boldsymbol{\varepsilon}(\omega,k)\cdot\mathbf{E}\end{aligned} \tag{A3}$$

at $\omega=\omega_{\mathrm{g}}$. Indeed, one could pick the following longitudinal-type solution: $(\mathbf{E},\mathbf{H})=(\mathbf{E}_0,0)$ and $\mathbf{k}=k_c\mathbf{E}_0/|\mathbf{E}_0|$. This contradicts the existence of a gap. Hence, $\det\boldsymbol{\varepsilon}(\omega,k)$ and $\det\boldsymbol{\mu}(\omega,k)$ cannot change sign in the gap. In particular, it follows that the number of positive (negative) eigenvalues of each tensor cannot change across the gap.

In general, the Maxwell eigenvalue problem in the considered system can be expressed as:

$$\mathbf{A}(\omega,\mathbf{k})\cdot\begin{pmatrix}\mathbf{E}\\ \mathbf{H}\end{pmatrix}=0 \qquad \text{with} \quad \mathbf{A}(\omega,\mathbf{k})=\begin{pmatrix}\omega\boldsymbol{\varepsilon}(\omega,k) & \mathbf{k}\times\mathbf{1}\\ -\mathbf{k}\times\mathbf{1} & \omega\boldsymbol{\mu}(\omega,k)\end{pmatrix}. \tag{A4}$$

A full gap requires that $\det\mathbf{A}(\omega_{\mathrm{g}},\mathbf{k})$ does not vanish for any real-valued $\mathbf{k}$. In particular, the number of positive (or negative) eigenvalues of $\mathbf{A}(\omega_{\mathrm{g}},\mathbf{k})$ cannot change as $\mathbf{k}$ is varied.

For large *k*, the matrix is asymptotically:

$$\mathbf{A}(\omega_{\mathrm{g}},\mathbf{k}\to\infty)\approx\begin{pmatrix}0 & \mathbf{k}\times\mathbf{1}\\ -\mathbf{k}\times\mathbf{1} & 0\end{pmatrix}, \tag{A5}$$

where it is implicit that the material parameters remain bounded for large $k$, as in Eq. (2). This matrix $\mathbf{A}\left(\omega_{\mathrm{g}}, \mathbf{k} \to \infty\right)$ can be diagonalized as:

$$\begin{aligned}\tilde{\mathbf{A}}\left(\omega_{\mathrm{g}}, \mathbf{k} \to \infty\right) = & -\frac{1}{2}\left[\begin{pmatrix}\mathbf{e}_{1\mathbf{k}} \\ \mathbf{e}_{2\mathbf{k}}\end{pmatrix} \otimes \begin{pmatrix}\mathbf{e}_{1\mathbf{k}} \\ \mathbf{e}_{2\mathbf{k}}\end{pmatrix} + \begin{pmatrix}\mathbf{e}_{2\mathbf{k}} \\ -\mathbf{e}_{1\mathbf{k}}\end{pmatrix} \otimes \begin{pmatrix}\mathbf{e}_{2\mathbf{k}} \\ -\mathbf{e}_{1\mathbf{k}}\end{pmatrix}\right] \\ & +\frac{1}{2}\left[\begin{pmatrix}\mathbf{e}_{1\mathbf{k}} \\ -\mathbf{e}_{2\mathbf{k}}\end{pmatrix} \otimes \begin{pmatrix}\mathbf{e}_{1\mathbf{k}} \\ -\mathbf{e}_{2\mathbf{k}}\end{pmatrix} + \begin{pmatrix}\mathbf{e}_{2\mathbf{k}} \\ \mathbf{e}_{1\mathbf{k}}\end{pmatrix} \otimes \begin{pmatrix}\mathbf{e}_{2\mathbf{k}} \\ \mathbf{e}_{1\mathbf{k}}\end{pmatrix}\right].\end{aligned} \tag{A6}$$

with $\mathbf{e}_{1\mathbf{k}}, \mathbf{e}_{2\mathbf{k}}, \hat{\mathbf{k}}$ a direct orthonormal basis with $\mathbf{e}_{1\mathbf{k}} \times \mathbf{e}_{2\mathbf{k}} = \hat{\mathbf{k}}$. The tilde indicates normalization of the matrix to $k$: $\tilde{\mathbf{A}} = \mathbf{A}/k$. The eigenvalues of $\tilde{\mathbf{A}}$ are −1, +1, and 0, all with double multiplicity.

For large but finite $k$, one can use degenerate perturbation theory to find how the eigenspace associated with the trivial eigenvalue (0) is perturbed. To this end, I write:

$$\tilde{\mathbf{A}}\left(\omega_{\mathrm{g}}, \mathbf{k}\right) = \tilde{\mathbf{A}}\left(\omega_{\mathrm{g}}, \mathbf{k} \to \infty\right) + \delta\tilde{\mathbf{A}}, \qquad \text{with } \delta\tilde{\mathbf{A}} = \frac{1}{k}\begin{pmatrix}\omega\boldsymbol{\varepsilon}(\omega, k) & 0 \\ 0 & \omega\boldsymbol{\mu}(\omega, k)\end{pmatrix}. \tag{A7}$$

Writing $\mathbf{f} = \mathbf{f}_{\infty} + \delta\mathbf{f}$, with $\mathbf{f}_{\infty} = \alpha_1 \mathbf{f}_{\infty 1} + \alpha_2 \mathbf{f}_{\infty 2}$ a linear combination of the two relevant degenerate eigenvectors, $\mathbf{f}_{\infty 1} = \begin{pmatrix}\hat{\mathbf{k}} \\ 0\end{pmatrix}$ and $\mathbf{f}_{\infty 2} = \begin{pmatrix}0 \\ \hat{\mathbf{k}}\end{pmatrix}$, one finds that to first order the perturbed eigenvalues ($\delta\lambda$) satisfy:

$$\left[\mathbf{f}_{\infty i}^{*} \cdot \delta\tilde{\mathbf{A}} \cdot \mathbf{f}_{\infty j}\right]\left[\alpha_j\right] = \delta\lambda\left[\alpha_j\right]. \tag{A8}$$

It is simple to see that the matrix is diagonal, and so the perturbed eigenvalues of $\tilde{\mathbf{A}}\left(\omega_{\mathrm{g}}, \mathbf{k}\right)$ are

$$\delta\lambda \approx \frac{1}{k}\omega\hat{\mathbf{k}} \cdot \boldsymbol{\varepsilon}\left(\omega, \infty\right) \cdot \hat{\mathbf{k}} \quad \text{and} \quad \delta\lambda \approx \frac{1}{k}\omega\hat{\mathbf{k}} \cdot \boldsymbol{\mu}\left(\omega, \infty\right) \cdot \hat{\mathbf{k}}. \tag{A9}$$

Restoring the normalization, the corresponding eigenvalues of $\mathbf{A}$ are asymptotically proportional to $\hat{\mathbf{k}} \cdot \boldsymbol{\varepsilon}\left(\omega, \infty\right) \cdot \hat{\mathbf{k}}$ and $\hat{\mathbf{k}} \cdot \boldsymbol{\mu}\left(\omega, \infty\right) \cdot \hat{\mathbf{k}}$, respectively. As already discussed, the eigenvalue branches cannot change sign as $\mathbf{k}$ is varied in the wave vector space. This shows that both the permittivity and the permeability are required to be either positive or negative definite tensors.

Furthermore, they cannot be both positive definite tensors. In fact, in that case the local asymptotic reference determined by $\boldsymbol{\varepsilon}_\infty(\omega)=\boldsymbol{\varepsilon}(\omega,\infty)$ and $\boldsymbol{\mu}_\infty(\omega)=\boldsymbol{\mu}(\omega,\infty)$ is described by two positive definite tensors. But it is well known that the reference material cannot be gapped in those circumstances: the isofrequency surfaces are closed Fresnel-type surfaces [55]. The case in which both tensors are negative definite is equivalent: defining $\boldsymbol{\varepsilon}'=-\boldsymbol{\varepsilon}$, $\boldsymbol{\mu}'=-\boldsymbol{\mu}$ and $\mathbf{H}'=-\mathbf{H}$ reduces Eq. (A3) to the positive-definite case.

In summary, the previous analysis shows that every reciprocal gapped system described by tensors of the form $\boldsymbol{\varepsilon}(\omega,k)$ and $\boldsymbol{\mu}(\omega,k)$ must, at any frequency $\omega$ within the gap, either be ENG-like for all *k* ($\boldsymbol{\varepsilon}(\omega,k)$ is negative definite for all *k* and $\boldsymbol{\mu}(\omega,k)$ is positive definite for all *k*) or MNG-like for all *k* ($\boldsymbol{\varepsilon}(\omega,k)$ is positive definite for all *k* and $\boldsymbol{\mu}(\omega,k)$ is negative definite for all *k*).

## Appendix B: Two-band model

This Appendix develops a two-band model to characterize the effect of a perturbation on the reciprocal uniaxial template of Sec. III.

For a perturbation $\delta\mathbf{M}$ of the material response as in Eq. (17), the Maxwell equations [Eq. (7)] may be approximated as

$$\left[\mathcal{D}(\mathbf{k})-\omega_c\delta\mathbf{M}(\omega_c,\mathbf{k})\right]\cdot\mathbf{f}_\mathbf{k}\approx\omega\mathbf{M}_0(\omega,\mathbf{k})\cdot\mathbf{f}_\mathbf{k}, \tag{B1}$$

where the dispersive effects in the perturbation are neglected. This is a frequency-dependent eigenvalue problem. To analyze its spectrum using perturbation theory, it is convenient to recast it as a standard eigenvalue problem.

This can be done using the methods of Refs. [7, 44], where the dispersion effects can be described through an extended state vector $\mathbf{Q}$ of the type $\mathbf{Q} = \begin{pmatrix} \mathbf{f} & \mathbf{Q}_1 & \mathbf{Q}_2 & ... \end{pmatrix}$. Here, the $\mathbf{Q}_\alpha$ model the internal degrees of freedom of the material. In this formalism, the frequency-dependent problem (B1) assumes a conventional form [7, 44]:

$$\hat{H}(\mathbf{k}) \cdot \mathbf{Q} = \omega \cdot \mathbf{Q}, \tag{B2}$$

with $\hat{H}(\mathbf{k}) = \hat{H}_0(\mathbf{k}) + \delta\hat{H}_{\text{int}}(\mathbf{k})$. Here, $\hat{H}_0(\mathbf{k})$ is the matrix operator that describes the augmented problem for the original reciprocal template [7], whereas $\delta\hat{H}_{\text{int}}(\mathbf{k})$ describes the perturbation. As the dispersive effects in the perturbation are neglected (see Eq. (B1)), the operator $\delta\hat{H}_{\text{int}}(\mathbf{k})$ acts only on the electromagnetic degrees of freedom ($\mathbf{f}$) of state vector $\mathbf{Q}$. It can be identified with the matrix $-\omega_c \delta\mathbf{M}(\omega_c, \mathbf{k})$.

Next, I construct an approximate solution for the eigensystem (B2) in the vicinity of the critical frequency $\omega_c$. To this end, write $\mathbf{Q} = \alpha_1 \mathbf{Q}_{\text{TE},\mathbf{k}} + \alpha_2 \mathbf{Q}_{\text{TM},\mathbf{k}}$, where $\mathbf{Q}_{\text{TE},\mathbf{k}}$ and $\mathbf{Q}_{\text{TM},\mathbf{k}}$ are the TE and TM eigenstates of the unperturbed (augmented) problem with $\omega_{\text{TE},\mathbf{k}} \approx \omega_c \approx \omega_{\text{TM},\mathbf{k}}$. Substituting this expansion into Eq. (B2), and projecting both members onto the two unperturbed eigenstates, one obtains the following two-band model:

$$\begin{pmatrix} \langle \mathbf{Q}_{\text{TE},\mathbf{k}} | \delta\hat{H}_{\text{int}} | \mathbf{Q}_{\text{TE},\mathbf{k}} \rangle & \langle \mathbf{Q}_{\text{TE},\mathbf{k}} | \delta\hat{H}_{\text{int}} | \mathbf{Q}_{\text{TM},\mathbf{k}} \rangle \\ \langle \mathbf{Q}_{\text{TM},\mathbf{k}} | \delta\hat{H}_{\text{int}} | \mathbf{Q}_{\text{TE},\mathbf{k}} \rangle & \langle \mathbf{Q}_{\text{TM},\mathbf{k}} | \delta\hat{H}_{\text{int}} | \mathbf{Q}_{\text{TM},\mathbf{k}} \rangle \end{pmatrix} \begin{pmatrix} \alpha_1 \\ \alpha_2 \end{pmatrix} = \begin{pmatrix} \omega - \omega_{\text{TE},\mathbf{k}} & 0 \\ 0 & \omega - \omega_{\text{TM},\mathbf{k}} \end{pmatrix} \begin{pmatrix} \alpha_1 \\ \alpha_2 \end{pmatrix}. \tag{B3}$$

Here, $\langle ... | ... \rangle$ can be identified with the canonical inner product in the augmented space [7, 44]. The eigenfunctions are normalized such that $\langle \mathbf{Q}_{i,\mathbf{k}} | \mathbf{Q}_{j,\mathbf{k}} \rangle = \delta_{ij}$ with *i,j* =TE,TM. The corresponding normalization for the electromagnetic components of $\mathbf{Q}$ may be chosen such that [7, 44]:

$$\mathbf{f}_{i,\mathbf{k}}^{*} \cdot \partial_{\omega} \left( \omega \mathbf{M}_{0} \right) \Big|_{\omega_{i,\mathbf{k}}} \cdot \mathbf{f}_{i,\mathbf{k}} = 1 . \tag{B4}$$

As already mentioned $\delta \hat{H}_{\text{int}}$ only acts on electromagnetic degrees of freedom ($\delta \hat{H}_{\text{int}} \to -\omega_c \delta \mathbf{M}$). Hence, the elements of the matrix on the left-hand side of Eq. (B3) can be expressed in terms of the electromagnetic modes: $\langle \mathbf{Q}_{i,\mathbf{k}} | \delta \hat{H}_{\text{int}} | \mathbf{Q}_{j,\mathbf{k}} \rangle = -\omega_c \mathbf{f}_{i,\mathbf{k}}^{*} \cdot \delta \mathbf{M} \cdot \mathbf{f}_{j,\mathbf{k}}$. This result leads to Eq. (18) of the main text, with the index 1 associated with the TE modes and the index 2 with the TM modes.

## Appendix C: Rotationally-symmetric gyrotropic systems

In this Appendix, it is shown that a nontrivial axion phase is incompatible with gyrotropic material responses invariant under continuous rotations about a fixed axis. Taking the symmetry axis along the *z*-direction, the electric and magnetic responses of the considered material class have the form:

$$\boldsymbol{\varepsilon}(\omega, k) = \begin{pmatrix} \varepsilon_t & i\varepsilon_g & 0 \\ -i\varepsilon_g & \varepsilon_t & 0 \\ 0 & 0 & \varepsilon_{zz} \end{pmatrix}, \qquad \boldsymbol{\mu}(\omega, k) = \begin{pmatrix} \mu_t & i\mu_g & 0 \\ -i\mu_g & \mu_t & 0 \\ 0 & 0 & \mu_{zz} \end{pmatrix}. \tag{C1}$$

It is implicit that all the matrix elements can vary with frequency and with the magnitude of the wave vector, in agreement with the material class represented by Eq. (11).

As shown in Appendix A, the existence of a gapped reciprocal reference requires that the material behaves as either ENG-like or MNG-like at $k=\infty$. For definiteness, the ENG-like case is assumed, so that $\text{pos}\left\{\boldsymbol{\varepsilon}\left(\omega_{\text{g}},k=\infty\right)\right\}=0$, with $\omega_{\text{g}}$ some frequency in a full gap. Then, as already discussed in Sec. III, having a nontrivial axion phase requires that $\text{pos}\left\{\boldsymbol{\varepsilon}\left(\omega_{\text{g}},k=0\right)\right\}=2$.

The eigenvalues of the permittivity matrix in Eq. (C1) are $\varepsilon_{\pm}\equiv\varepsilon_{t}\pm\varepsilon_{g}$ and $\varepsilon_{zz}$. For a nontrivial axion phase, one needs that exactly two of these numbers are positive at the origin $k=0$, and that $\varepsilon_{t},\varepsilon_{zz}$ are both negative at $k=\infty$.

Importantly, $\varepsilon_{zz}$ and $\mu_{zz}$ are not allowed to change sign in the gap, because at a zero one would have a longitudinal mode. The argument is the same as in Appendix A. For example, if $\varepsilon_{zz}\left(\omega_{\text{g}},k_{c}\right)=0$ for some real-valued $k_{c}$, then for $\mathbf{k}=k_{c}\hat{\mathbf{z}}$, Maxwell's equations would admit the solution $\mathbf{E}=E_{0}\hat{\mathbf{z}},\mathbf{H}=0$ (propagation factor was omitted), which would contradict the existence of a full gap. Thus, considering the ENG-like character at infinity, it follows that the existence of a full gap requires $\varepsilon_{zz}\left(\omega_{\text{g}},k\right)<0$ and $\mu_{zz}\left(\omega_{\text{g}},k\right)>0$ for all *k*. Moreover, this result implies that a nontrivial axion phase is feasible only if $\varepsilon_{+}\left(\omega_{\text{g}},k=0\right)$ and $\varepsilon_{-}\left(\omega_{\text{g}},k=0\right)$ are both positive.

Next, it is demonstrated that $\varepsilon_{\pm}\left(\omega_{\text{g}},k=0\right)>0$, $\varepsilon_{t}\left(\omega_{\text{g}},k=\infty\right)<0$ and $\mu_{zz}\left(\omega_{\text{g}},k\right)>0$ are incompatible with having a full gap. It is enough to consider propagation in the $k_{z}=0$ plane and TE-polarized waves. The dispersion of these waves is determined by:

$$\underbrace{\varepsilon_{t}k_{t}^{2}-\omega_{g}^{2}\varepsilon_{+}\varepsilon_{-}\mu_{zz}}_{f(k_{t})}=0. \tag{C2}$$

Evidently, the function $f(k_t)$ is negative for $k_t \to \infty$, assuming a response consistent with the structure in Eq. (11) and the ENG-like behavior at infinity. Hence, the system can be gapped only if $f(k_t) < 0$ for all $k_t$.

However, from the previous considerations the functions $\varepsilon_{\pm}(\omega_{\mathrm{g}}, k_t)$ are required to have opposite signs at the two endpoints (note that $\varepsilon_{\pm}(\omega_{\mathrm{g}}, \infty) = \varepsilon_t(\omega_{\mathrm{g}}, \infty) < 0$), and thereby they must cross zero at some wave number. Let $k_c$ be the smallest of the zeros of $\varepsilon_{+}(\omega_{\mathrm{g}}, k_t)$ and $\varepsilon_{-}(\omega_{\mathrm{g}}, k_t)$. The condition $f(k_t) < 0$ requires that $\varepsilon_t(\omega_{\mathrm{g}}, k_c)$ is negative. But in these conditions, one of two circular permittivities, $\varepsilon_{\pm} \equiv \varepsilon_t \pm \varepsilon_g$, must also be negative at the same crossing $k_t = k_c$. By construction, $\varepsilon_{\pm}$ cannot change sign in the interval $0 < k_t < k_c$, and hence it follows that $\varepsilon_{\pm}$ cannot both be positive at the origin $k_t = 0$. This is a contradiction, and thus these systems cannot host an axion phase.

## Appendix D: Magnetoelectric perturbation

In this Appendix, I apply the first-order gap criterion in Eq. (23) to an odd magnetoelectric perturbation of the reciprocal template of the form:

$$\delta\mathbf{M} = \begin{pmatrix} 0 & \boldsymbol{\xi}(\omega,\mathbf{k}) \\ \left[\boldsymbol{\xi}(\omega,\mathbf{k})\right]^{\dagger} & 0 \end{pmatrix}. \tag{D1}$$

For this perturbation, Eq. (23) reduces to:

$$\mathbf{E}_{\mathbf{k}1}^{*} \cdot \boldsymbol{\xi}(\omega_{\mathrm{c}},\mathbf{k}) \cdot \mathbf{H}_{\mathbf{k}2} + \mathbf{H}_{\mathbf{k}1}^{*} \cdot \left[\boldsymbol{\xi}(\omega_{\mathrm{c}},\mathbf{k})\right]^{\dagger} \cdot \mathbf{E}_{\mathbf{k}2} \neq 0, \tag{D2}$$

on the toroidal surface. As before, the subscripts 1 and 2 label the electric and magnetic fields of the TE- and TM- polarized eigenstates, respectively. It is useful to note that for the reciprocal template the fields $\mathbf{f}_\mathbf{k} = \left(\mathbf{E}_\mathbf{k} \quad \mathbf{H}_\mathbf{k}\right)^T$ associated with the TE and TM eigenstates are

$$\mathbf{f}_{\mathrm{TE},\mathbf{k}} \sim \begin{pmatrix} \hat{\boldsymbol{\varphi}} \\ -\dfrac{k_z}{\omega_c \mu_t}\hat{\boldsymbol{\rho}} + \dfrac{k_t}{\omega_c \mu_{zz}}\hat{\mathbf{z}} \end{pmatrix}, \qquad \mathbf{f}_{\mathrm{TM},\mathbf{k}} \sim \begin{pmatrix} \dfrac{k_z}{\omega_c \varepsilon_t}\hat{\boldsymbol{\rho}} - \dfrac{k_t}{\omega_c \varepsilon_{zz}}\hat{\mathbf{z}} \\ \hat{\boldsymbol{\varphi}} \end{pmatrix}, \tag{D3}$$

where I used $\mathbf{k} = k_t \hat{\boldsymbol{\rho}} + k_z \hat{\mathbf{z}}$ on the toroidal isofrequency surface. Here, $\hat{\boldsymbol{\rho}}, \hat{\boldsymbol{\varphi}}, \hat{\mathbf{z}}$ are the unit vectors associated with a system of cylindrical coordinates in momentum space.

When the coupling is reciprocal, $\boldsymbol{\xi}(\omega,\mathbf{k})$ is real-valued, and the condition (D2) is violated on the toroidal surface. In fact, considering the same outer longitudinal-mode circle as in Sec IV B, Eq. (D2) reduces to $\hat{\boldsymbol{\rho}} \cdot \boldsymbol{\xi}(\omega_c, k_L \hat{\boldsymbol{\rho}}) \cdot \hat{\boldsymbol{\rho}} \neq 0$, where $k_L$ is the outer circle radius. However, for an odd magnetoelectric response $\hat{\boldsymbol{\rho}} \cdot \boldsymbol{\xi}(\omega_c, k_L \hat{\boldsymbol{\rho}}) \cdot \hat{\boldsymbol{\rho}}$ changes sign under the transformation $\hat{\boldsymbol{\rho}} \to -\hat{\boldsymbol{\rho}}$. Thus, it must have a root in the considered outer circle. Note that the same obstruction arises when the coupling is purely imaginary.

For a complex-valued coupling ($\boldsymbol{\xi} = \boldsymbol{\xi}_\mathrm{R} + i\boldsymbol{\xi}_\mathrm{I}$, with $\boldsymbol{\xi}_\mathrm{R}$ and $\boldsymbol{\xi}_\mathrm{I}$ real-valued tensors) the situation changes qualitatively. In that case, Eq. (D2) splits into two independent conditions, resulting in the following constraint:

$$\mathbf{E}_{\mathbf{k}1} \cdot \boldsymbol{\xi}_\mathrm{R} \cdot \mathbf{H}_{\mathbf{k}2} + \mathbf{E}_{\mathbf{k}2} \cdot \boldsymbol{\xi}_\mathrm{R} \cdot \mathbf{H}_{\mathbf{k}1} \quad \text{and} \quad \mathbf{E}_{\mathbf{k}1} \cdot \boldsymbol{\xi}_\mathrm{I} \cdot \mathbf{H}_{\mathbf{k}2} - \mathbf{E}_{\mathbf{k}2} \cdot \boldsymbol{\xi}_\mathrm{I} \cdot \mathbf{H}_{\mathbf{k}1}$$

cannot vanish *simultaneously* on the toroidal surface. (D4)

I took into account that the eigenstates of the reciprocal template can be chosen real-valued. Hence, for a complex-valued magnetoelectric coupling, the first-order direct gap can fail to open only when the above two real functions have coincident zeros on the toroidal surface.

It is interesting to note that a magnetoelectric coupling can always be formally absorbed into an effective Landau–Lifshitz permittivity defined as [56, 57]:

$$\boldsymbol{\varepsilon}_{\mathrm{LL}}(\omega,\mathbf{k})=\boldsymbol{\varepsilon}-\boldsymbol{\xi}\cdot\boldsymbol{\mu}^{-1}\cdot\boldsymbol{\zeta}+\left[\boldsymbol{\xi}\cdot\boldsymbol{\mu}^{-1}\times\frac{\mathbf{k}}{\omega}-\frac{\mathbf{k}}{\omega}\times\left(\boldsymbol{\mu}^{-1}\cdot\boldsymbol{\zeta}\right)\right]+\frac{\mathbf{k}}{\omega}\times\left(\boldsymbol{\mu}^{-1}-\mu_0^{-1}\mathbf{1}\right)\times\frac{\mathbf{k}}{\omega}. \qquad \text{(D5)}$$

It was shown in Sec. IV B that purely electric perturbations of the reciprocal template cannot gap it within first order perturbation theory. Thus, it may appear contradictory that a magnetoelectric coupling perturbation (e.g., the perturbation in Eq. (26)) can gap the reciprocal template, given that it can always be expressed as an equivalent $\boldsymbol{\varepsilon}_{\mathrm{LL}}$ permittivity tensor, as shown above. The resolution is that the effective permittivity generated by the Landau–Lifshitz transformation is not a weak perturbation of the reciprocal template. Indeed, for the reciprocal template, the transformation involves the inverse permeability, which diverges on the outer circle of the $k_z=0$ cut of the toroidal surface because $\mu_t=0$ there.